\documentstyle[]{mn}

\newif\ifAMStwofonts

\input psfig.sty

\ifoldfss
  \newcommand{\rmn}[1] {{\rm #1}}

  \ifCUPmtlplainloaded \else
    \NewTextAlphabet{textbfit} {cmbxti10} {}
    \NewTextAlphabet{textbfss} {cmssbx10} {}
    \NewMathAlphabet{mathbfit} {cmbxti10} {} 
    \NewMathAlphabet{mathbfss} {cmssbx10} {} 
  \fi
  \ifAMStwofonts
    \ifCUPmtlplainloaded \else
      \NewSymbolFont{upmath} {eurm10}
      \NewSymbolFont{AMSa} {msam10}
      \NewMathSymbol{\upi}     {0}{upmath}{19}
      \NewMathSymbol{\umu}     {0}{upmath}{16}
      \NewMathSymbol{\upartial}{0}{upmath}{40}
      \NewMathSymbol{\leqslant}{3}{AMSa}{36}
      \NewMathSymbol{\geqslant}{3}{AMSa}{3E}

      \let\leq=\leqslant 
      \let\geq=\geqslant 
    \fi
  \fi
\fi 

\ifnfssone
  \newmathalphabet{\mathit}
  \addtoversion{normal}{\mathit}{cmr}{m}{it}
  \addtoversion{bold}{\mathit}{cmr}{bx}{it}
  \newcommand{\rmn}[1] {\mathrm{#1}}

  \newmathalphabet{\mathbfit} 
  \addtoversion{normal}{\mathbfit}{cmr}{bx}{it}
  \addtoversion{bold}{\mathbfit}{cmr}{bx}{it}
  \newmathalphabet{\mathbfss} 
  \addtoversion{normal}{\mathbfss}{cmss}{bx}{n}
  \addtoversion{bold}{\mathbfss}{cmss}{bx}{n}
  \ifAMStwofonts
    \ifCUPmtlplainloaded \else
      \UseAMStwoboldmath
      \makeatletter
      \new@mathgroup\upmath@group
      \define@mathgroup\mv@normal\upmath@group{eur}{m}{n}
      \define@mathgroup\mv@bold\upmath@group{eur}{b}{n}
      \edef\UPM{\hexnumber\upmath@group}
      \new@mathgroup\amsa@group
      \define@mathgroup\mv@normal\amsa@group{msa}{m}{n}
      \define@mathgroup\mv@bold\amsa@group{msa}{m}{n}
      \edef\AMSa{\hexnumber\amsa@group}
      \makeatother
      \mathchardef\upi="0\UPM19
      \mathchardef\umu="0\UPM16
      \mathchardef\upartial="0\UPM40
      \mathchardef\leqslant="3\AMSa36
      \mathchardef\geqslant="3\AMSa3E

      \let\leq=\leqslant 
      \let\geq=\geqslant 
    \fi
  \fi
\fi 

\ifnfsstwo
  \newcommand{\rmn}[1] {\mathrm{#1}}

  \DeclareMathAlphabet{\mathbfit}{OT1}{cmr}{bx}{it}
  \SetMathAlphabet\mathbfit{bold}{OT1}{cmr}{bx}{it}
  \DeclareMathAlphabet{\mathbfss}{OT1}{cmss}{bx}{n}
  \SetMathAlphabet\mathbfss{bold}{OT1}{cmss}{bx}{n}
  \ifAMStwofonts
    \ifCUPmtlplainloaded \else
      \DeclareSymbolFont{UPM}{U}{eur}{m}{n}
      \SetSymbolFont{UPM}{bold}{U}{eur}{b}{n}
      \DeclareSymbolFont{AMSa}{U}{msa}{m}{n}
      \DeclareMathSymbol{\upi}{0}{UPM}{"19}
      \DeclareMathSymbol{\umu}{0}{UPM}{"16}
      \DeclareMathSymbol{\upartial}{0}{UPM}{"40}
      \DeclareMathSymbol{\leqslant}{3}{AMSa}{"36}
      \DeclareMathSymbol{\geqslant}{3}{AMSa}{"3E}

      \let\leq=\leqslant 
      \let\geq=\geqslant 
    \fi
  \fi
\fi 

\ifCUPmtlplainloaded \else
  \ifAMStwofonts \else 
    \def\upi{\pi}
    \def\umu{\mu}
    \def\upartial{\partial}
  \fi
\fi

\title[The Clustering of Colour Selected Galaxies]
{The Clustering of Colour Selected Galaxies}
\author[M.~J.~I.~Brown, R.~L.~Webster and B.~J.~Boyle]
       {M.~J.~I.~Brown,$^1$\thanks{Email:~mbrown@physics.unimelb.edu.au} 
R.~L.~Webster$^1$ and B.~J.~Boyle$^2$ \\
$^1$School of Physics, University of Melbourne, Parkville, Victoria 3052, Australia\\
$^2$Anglo-Australian Observatory, P.O. Box 296, Epping, NSW 1710, Australia}
\date{Accepted 19?? ???????? ??.
      Received 19?? ???????? ??;
      in original form 19?? ???????? ??}

\pagerange{\pageref{firstpage}--\pageref{lastpage}}
\pubyear{1999}

\begin{document}

\maketitle

\label{firstpage}

\begin{abstract}
We present measurements of the angular correlation function 
of galaxies selected from a $B_J\sim 23.5$ multicolour survey of two 
$5^\circ \times 5^\circ$ fields located at high galactic latitudes. 
The galaxy catalogue of $\sim 4\times 10^5$ galaxies is comparable 
in size to catalogues used
to determine the galaxy correlation function at low-redshift. 
Measurements of the $z\sim 0.4$ correlation function 
at large angular scales show
no evidence for a break from a power law  though our results
are not inconsistent with a break at $\ga 15 h^{-1} {\rmn Mpc}$. 
Despite the large fields-of-view, there are large discrepancies between 
the measurements of the correlation function in each field, possibly due to 
dwarf galaxies within $z\sim 0.11$ clusters near the South Galactic Pole. 

Colour selection is used to study the clustering of galaxies
$z\sim 0$ to $z\sim 0.4$. The galaxy correlation function
is found to strongly depend on colour with red galaxies 
more strongly clustered than blue galaxies by a factor 
of $\ga 5$ at small scales. The slope of the correlation
function is also found to vary with colour with $\gamma \sim 1.8$ for 
red galaxies while $\gamma \sim 1.5$ for blue galaxies. 
The clustering of red galaxies is consistently strong over the entire
magnitude range studied though there are large variations between the 
two fields. The clustering of blue galaxies is extremely weak over
the observed magnitude range with clustering consistent with 
$r_0\sim 2 h^{-1} {\rmn Mpc}$. This is weaker than the clustering
of late-type galaxies in the local Universe and suggests galaxy 
clustering is more strongly correlated with colour than 
morphology. This may also be the first detection of a substantial 
low redshift galaxy population with clustering properties similar to faint
blue galaxies. 
\end{abstract}

\begin{keywords}
(cosmology:) large-scale structure of Universe  -- galaxies: evolution.
\end{keywords}

\section{Introduction}

The galaxy two-point correlation function is commonly used to 
measure the structure of the galaxy environment from high redshift until the 
present epoch. The clustering properties of galaxies in the local
Universe are well measured by large representative surveys of 
the galaxy population (Maddox, Efstathiou \& Sutherland 1996).
Catalogues of galaxies selected by morphology show large variations of 
the galaxy correlation function 
with late type galaxies having considerably weaker clustering
than early type galaxies (Davis \& Geller 1976, Loveday {\it et al.} 1995). 

The results from studies of galaxies with fainter apparent
magnitudes and higher redshifts are less conclusive. Pencil-beam surveys 
with CCDs and  photographic plates from $4{\rmn m}$ telescopes have
measured the amplitude of the $B>22$ correlation function;
however, estimates vary by $\ga 100\%$ (Infante \& Pritchet 1995). 
Also, while $B>22$ surveys show evidence for a rapid decline of 
the amplitude of the correlation function (Efstathiou {\it et al.} 1991, 
Infante \& Pritchet 1995, Roche {\it et al.} 1996), $I$ band imaging surveys 
to similar depths show no evidence for a rapid decrease of the correlation
function amplitude (Postman {\it et al.} 1998). 

The small areas of previous studies of the faint galaxy correlation
function are a possible source of the discrepancy. 
Large individual structures and voids in the Universe could bias
estimates of the correlation function if the field-of-view of the survey
is small. The use of single band data to select catalogues of
galaxies could suffer from biases as the morphological mix
of galaxies will change as a function of limiting magnitude. 
It is probable that the differing amplitudes of the $B$ and $I$ band
correlation functions are due to faint $B$ band data being dominated
by weakly clustered blue galaxies (Efstathiou {\it et al.} 1991)
while the $I$ band data has a larger fraction of early type galaxies. 

In this paper, we use a $B_J\sim 23.5$ multicolour catalogue 
of galaxies derived from two $5^\circ \times 5^\circ$ fields 
to measure the clustering properties of faint galaxies. 
Section 2 discusses the observations and data reduction used to produce
the galaxy catalogue. The method used to determine the angular
and spatial correlation functions is described in Sections 3 and 4. 
Estimates of the correlation function at large angular scales are
presented in Section 5. Section 6 discusses the angular correlation 
function as a function of limiting magnitude for the $U$, $B_J$, $R_F$ and 
$I$ bands. We show that it is impossible to use a single model to describe the 
observed clustering of galaxies as different populations of galaxies
are measured by each band as a function of limiting magnitude. In Section 7, we
use colour selection to measure galaxies with similar stellar populations over
a range of redshifts. Our estimates of the spatial correlation function 
indicate clustering is more strongly correlated with colour than morphology. 
Our main conclusions are summarised in Section 8. 

\section{The Galaxy Catalogue}

The image data consists of $U$, $B_J$, $R_F$ and $I$ band digitally
coadded SuperCOSMOS scans of photographic plates 
of the South Galactic Pole Field (SGP) and 
UK Schmidt field 855 (F855). The field centres, number of photographic plates 
and limiting magnitudes for each field and band are listed in 
Table~\ref{table:stacks}. 
The individual photographic plates were obtained with the 
$1.25{\rmn{m}}$ UK Schmidt Telescope between 1977 and 1998
for various programmes. 
The field-of-view is $6.4^\circ \times 6.4^\circ$ 
though vignetting is greater than $0.1$ magnitudes at angular scales 
more than $\sim 3.35^\circ$ from the optical axis (Tritton 1983). 

The individual plates were digitised with the SuperCOSMOS 
facility at Royal Observatory Edinburgh (Miller {\it et al.} 1991). The 
$10 \mu {\rmn{m}}$ pixels and $15 \mu{\rmn{m}}$ resolution 
of SuperCOSMOS corresponds to a pixel scale of $0.67^{\prime \prime}$ and 
allows good sampling of images obtained in 
typical seeing ($\sim 2.5^{\prime \prime}$).
The plate scans were converted
from transmission to approximate intensity space before stacking. 
All plates were sky-limited so faint objects are in the linear regime
of the plate response. 

Before coadding the individual plate scans, the background
was subtracted with a $160 \times 160$ pixel median filter. 
Without background subtraction, robust bad pixel rejection 
is not possible resulting in contamination of the catalogue. 
To reduce the computation time required to subtract the 
background, the background was calculated for $32 \times 32$ pixel 
regions rather than for each pixel and only every 4th pixel in the
$160 \times 160$ filter box was used to determine the median. 
Before coadding the scans, the noise distribution of the
individual plates was measured and found to be fitted accurately by 
a Gaussian curve. If the noise distribution was significantly skewed, large
systematic errors would propagate into the coadded plate scans. 

As faint objects are in the linear regime of the plate response and
the noise distribution is approximately Gaussian, the 
IRAF\footnote{IRAF 
is distributed by the National Optical Astronomy Observatories,
which are operated by the Association of Universities for Research
in Astronomy, Inc., under cooperative agreement with the National
Science Foundation.}
task {\tt imcombine} can be effectively used to coadd the data. 
An average of the plates weighted by the background noise and 
min-max bad pixel rejection is used to provide high signal-to-noise and 
robust bad pixel rejection.
Without robust bad pixel rejection, there is significant contamination of the 
catalogue by plate-flaws, satellite trails and asteroids. For bands
with 5 or less plates, the number of pixels rejected by the bad
pixel rejection algorithm results in the coaddition effectively being
a median. For the SGP $U$ band data, where only 2 of the 3 plates are 
of high quality, a weighted average with no bad pixel rejection is 
used and objects are correlated with the $B_J$, $R_F$ and $I$ catalogues
to remove contamination. The $U$ band data is complete to $U=22$ for 
objects with $B_J$ detections but incompleteness does effect the 
catalogue at $U>21.5$ for faint galaxies with  $U-B_J<-1.5$. 

Object detection, instrumental photometry 
and faint object star-galaxy classifications are determined using
SExtractor (Bertin \& Arnouts 1996).
In the non-linear regime of the plate response where SExtractor 
classifications are unreliable, 
objects are classified as stellar if they are within 
the stellar locus in peak intensity versus magnitude 
space and area versus magnitude space. 
The reliability of the object classifications is improved by using 
object classifications in multiple bands where possible. 
Objects with SExtractor classification scores greater than 
0.7 in 3 bands, 0.75 in 2 bands or 0.85 in a single band are classified
as stellar objects. Comparison of the object 
classification with the $B_J<22$ and $B<24$ spectroscopic 
samples of Colless {\it et al.} (1990) and Glazebrook {\it et al.} (1995)
show the star-galaxy separation to be correct for $\sim 85 \%$ of $B_J<23$ 
objects and contamination of the galaxy catalogue is $\la 5\%$. 
At $B_J<20$, comparison of 
object classifications with CCD images, eye-ball classifications 
and colour-colour diagrams indicates 
the reliability of the star-galaxy separation is $\ga 95 \%$. 

Photometric calibration of the data was obtained with 
$B$, $V$, $R$ and $I$ band CCD data from the Siding Spring 40-inch
provided by Bruce Peterson, $B$ and $R$ band CCD imaging with the 
Anglo-Australian Telescope provided by Bryn Jones and $U$ 
band photometry from Croom {\it et al.} (1999) and Osmer {\it et al.} (1998). 
Galaxy and faint stellar photometry is obtained by adding a 
zero-point value to the instrumental magnitudes while bright stellar 
photometry is determined by a polynomial
fit between the instrumental magnitudes and calibrated CCD magnitudes.
All magnitudes are corrected for vignetting using a fit to the 
vignetting as a function of radius plot from Tritton (1983). 
Comparison of the photometry with Boyle, Shanks \& Croom (1995), 
Caldwell \& Schechter (1996), Croom {\it et al.} (1999), 
Osmer {\it et al.} (1998) and Patch B of the ESO 
Imaging Survey (Prandoni {\it et al.} 1999) limits zero point 
errors to less than $0.05$ magnitudes. Comparison of 
median colours as a function of limiting magnitude and 
galaxy loci in colour-colour diagrams 
show no significant offsets between the two fields. 

\begin{figure}
\psfig{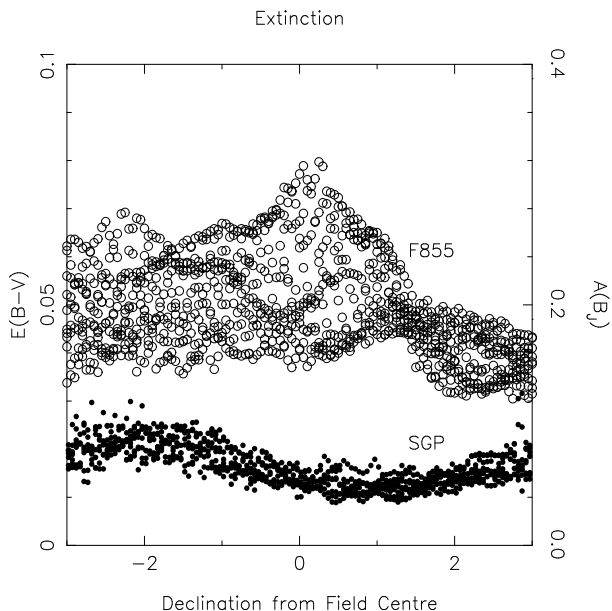}
\caption{A plot of the dust extinction across the SGP and F855 fields
using $E(B-V)$ estimates from Schlegel, Finkbeiner \& Davis (1998).
The dust extinction in F855 ($l\sim 45^{\circ}$) is significantly
larger and shows more structure than the SGP. The extinction in F855, if
left uncorrected, would introduce spurious structure on large
angular scales.}
\label{figure:dust}
\end{figure}

Magnitude estimates for galaxies are corrected for dust extinction
using the dust maps of Schlegel, Finkbeiner \& Davis (1998). 
Figure~\ref{figure:dust} shows that there are significant variations
in the dust extinction in F855 which could introduce spurious 
large-scale structure into the catalogue. In contrast, the dust 
extinction in the SGP is restricted to the range $0.01<E(B-V)<0.03$ 
which is a comparable to the $0.028$ magnitude error estimate of 
the dust maps. We therefore use the dust map estimates of the 
extinction for F855 while using a constant value of $E(B-V)=0.015$ 
to correct for dust extinction in the SGP. 

\begin{figure}
\psfig{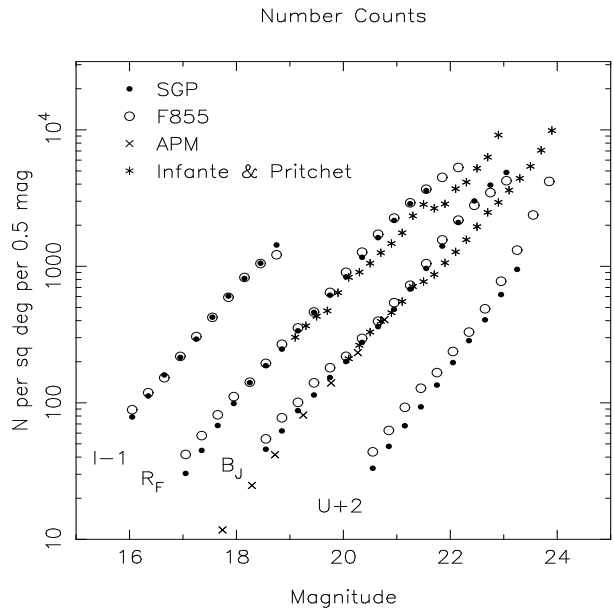}
\caption{A plot of the galaxy number counts as a function of magnitude
(after correction for dust extinction). 
Data from both fields is compared with number counts from 
the APM survey (Maddox {\it et al.} 1990)
and Infante \& Pritchet (1992).
$I$ band counts have been moved 1 magnitude to the left and 
$U$ band counts have been moved 2 magnitudes to the right for clarity.}
\label{figure:counts}
\end{figure}

Completeness limits for the catalogue are determined with comparison to
deeper CCD images (where available) and number counts as a function 
of limiting magnitude (Figure~\ref{figure:counts}). 
The number counts for the two fields are
in good agreement with each other in all 4 bands. However, at 
$B_J<19$ and $B_J>21$ the number counts are slightly higher
than those measured by Maddox {\it et al.} (1990) and 
Infante \& Pritchet (1992). 
The resulting completeness limits for the four bands are 
$U=21.5$, $B_J=23.5$, $R_F=22$ and $I=20$ for the SGP and 
$U=22$, $B_J=23.5$, $R_F=22.5$ and $I=20$ for F855. 
Overestimates of the completeness limits would result in biases in
estimates of the correlation function as vignetting would introduce
spurious structure into the catalogue. 

\begin{table}
\caption{Image data parameters}
\label{table:stacks}
\begin{tabular}{llccc}
Field & RA (1950) DEC 	& Band 		& Limiting	& \# of 	\\
      & 	        &  	        & Mag 	& Plates    \\ 
\\
SGP & $00^{\rmn{h}}~53^{\rmn{m}}$ $-28^\circ~03'$ & $U$	  & 21.5 & 3 \\
SGP & $00^{\rmn{h}}~53^{\rmn{m}}$ $-28^\circ~03'$ & $B_J$ & 23.5 & 9 \\
SGP & $00^{\rmn{h}}~53^{\rmn{m}}$ $-28^\circ~03'$ & $R_F$ & 22	& 8 \\
SGP & $00^{\rmn{h}}~53^{\rmn{m}}$ $-28^\circ~03'$ & $I$	  & 20	& 5 \\
F855 & $10^{\rmn{h}}~40^{\rmn{m}}$ $00^\circ~00'$ & $U$	  & 22 & 5 \\
F855 & $10^{\rmn{h}}~40^{\rmn{m}}$ $00^\circ~00'$ & $B_J$ & 23.5 & 12	\\
F855 & $10^{\rmn{h}}~40^{\rmn{m}}$ $00^\circ~00'$ & $R_F$ & 22.5 & 26 \\
F855 &	$10^{\rmn{h}}~40^{\rmn{m}}$ $00^\circ~00'$ & $I$  & 20 & 7 \\
\\
\end{tabular}
\end{table}

\begin{figure}
\psfig{file=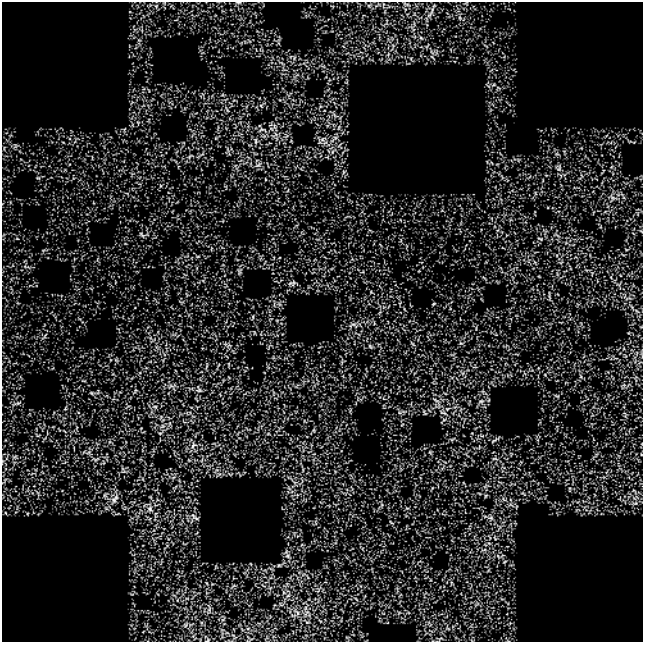,width=85mm,height=85mm,angle=0}
\caption{A flux weighted map of $R_F<22$ galaxies in the SGP. 
The dark blocks
are regions which have been removed from the catalogue. 
The corners of both fields have been removed to prevent objects with 
significant vignetting entering the catalogue.
Some of the Abell clusters in the field 
can be seen as overdense regions.}
\label{figure:map}
\end{figure}

Contamination of the catalogue by spurious objects biases estimates
of the galaxy correlation function and these have to be removed. 
Most plate-flaws, asteroids and satellite trails 
are effectively removed by the bad-pixel rejection algorithm. 
The plate edges, the step wedges, globular clusters, bright galaxies and the 
internal reflections of $V<5$ stars are manually removed from the catalogue. 
The majority of the remaining contaminating objects are diffraction spikes and 
halos around bright stars and these are removed with a program that 
drills regions around bright stars. Spurious structure may also be introduced
into the catalogue at angular scales more than $3.35^\circ$ from the plate
centre where number counts are altered by $>10\%$ during
the vignetting correction. To prevent this, the corners of the fields
have been excluded from the catalogue. The resulting geometry of the 
$5^\circ \times 5^\circ$ field-of-view is shown in Figure~\ref{figure:map}. 
The final catalogues of galaxies contain $\sim 2\times 10^5$
galaxies per field in both $B_J$ and $R_F$.

\section{Estimation of the Angular Correlation Function}

The galaxy two-point correlation function, $\omega(\theta)$, 
measures the mean excess surface density of pairs at angular separation 
$\theta$ compared with the expected number of pairs if 
galaxies were randomly distributed.
The most
commonly used estimator of the angular correlation function is 
\begin{equation}
\hat \omega (\theta) = \frac{DD}{DR} -1 
\end{equation}
where $DD$ and $DR$ are the number of galaxy-galaxy and galaxy-random 
object pairs at angular separations $\theta \pm \delta \theta$. The 
random objects are typically copies of real objects distributed 
randomly across the field-of-view. To reduce errors, multiple random
copies of each object can be made and the estimate of $DR$ renormalised. 
However, the $DD/DR$ estimator is subject
to first-order errors in the galaxy density contrast (Hamilton 1993) making
it unsuitable for measuring weak clustering at large angular scales. 
We therefore use an estimator with lower variance than $DD/DR$, the estimator 
\begin{equation}
\hat \omega (\theta) = \frac{DD - 2 DR + RR}{RR}
\end{equation}
(Landy \& Szalay 1993) where $RR$ is the number of random-random 
object pairs at angular separations $\theta \pm \delta \theta$.  
The value of $\hat \omega (\theta)$
is determined for different angular scale bins with the values of 
$DD$, $DR$ and $RR$ being determined with pairs of individual objects at
small angular scales and weighted pairs of cells containing multiple objects
at large angular scales. The use of cells to determine the angular 
correlation function reduces the computational time required 
to several hours and does not introduce significant errors. 

The estimator of the angular correlation function satisfies 
the integral constraint, 
\begin{equation}
\int \int \hat \omega (\theta) \delta \Omega_1 \delta \Omega_2 \simeq 0 
\end{equation}
(Groth and Peebles 1977), resulting in an underestimate of the angular correlation function. 
To remove this bias from the correlation function, the term 
\begin{equation}
\omega(\theta)_\Omega = 
\frac{1}{\Omega^2} \int \int \omega (\theta) \delta \Omega_1 \delta \Omega_2
\end{equation}
is added to the estimate of the correlation function. The term, 
$\omega (\theta)_\Omega$ does require an assumption of the form of the 
correlation function to correctly estimate the value of correlation 
function. However, previous work with smaller fields of view and
at brighter limiting magnitudes shows that the angular correlation 
is well approximated by a power law at angular scales less than 
$1^\circ$ (Maddox, Efstathiou \& Sutherland 1996). 

A further source of bias in estimates of the correlation function is
contamination of the catalogue by randomly distributed objects 
such as stars. If the fraction of the catalogue contaminated by stars 
is $f$, the estimate of the correlation function is reduced by a 
factor of $(1-f)^2$ at all angular separations. As mentioned previously, 
comparison with spectroscopic samples indicates the star galaxy 
separations if $\ga 85\%$ reliable at magnitude brighter than 
$B_J<22$. At fainter magnitudes, the contamination of the catalogue
by stars is negligible as galaxies consist more than $\ga 80\% $ of 
the object number counts at $B_J\geq 23$ (Glazebrook {\it et al.} 1995). 

\section{Modelling the Spatial Correlation Function}

For this work, we assume that the spatial correlation function is 
a power law of the form,
\begin{equation}
\xi(r,z) = \left( \frac{r}{r_0} \right) ^{-\gamma} (1+z)^{-(3+\epsilon)},
\end{equation}
(Efstathiou {\it et al.} 1991) 
where $r$ is the spatial separation in physical coordinates, $z$ is the 
redshift and $r_0$, $\gamma$ and $\epsilon$ are constants. If  
$\epsilon=0$ the clustering is fixed in physical coordinates while
if $\epsilon = \gamma - 3$ the clustering is fixed in comoving coordinates. 
This parameterisation of the evolution of the spatial correlation 
function is not valid at all redshifts but
is a good approximation at $z<1$ (Baugh {\it et al.} 1999). 
For galaxies selected with images in a single broadband, typical
values of the parameters of $\xi(r,z)$ are 
$r_0 \sim 5 h^{-1} {\rmn{Mpc}}$, $\gamma \sim 1.7$
(Maddox, Efstathiou \& Sutherland 1996) and $\epsilon \sim -1$. 

For a power law spatial correlation function, the resulting angular 
correlation function is a power law with 
\begin{equation}
\omega(\theta) = \sqrt{\pi}
\frac{\Gamma[(\gamma - 1)/2]}{\Gamma(\gamma/2)}B
r_0^{\gamma} \theta ^{(1-\gamma)}
\end{equation}
(Baugh \& Efstathiou 1993) 
where $B$ is a constant. The value of $B$ is given by 
\begin{equation}
B = \int^{\infty}_{0} g(z) \left(\frac{dN(z)}{dz}\right)^2 dz \left/ 
\left[ \int^{\infty}_{0} \frac {dN(z)}{dz} dz  \right]^{2} \right.
\end{equation}
where 
\begin{equation}
g(z) = \frac{dz}{dx}x^{1-\gamma}F(x)(1+z)^{-(3+\epsilon-\gamma)},
\end{equation}
$x$ is the coordinate distance at redshift z, $dN/dz$ is the number of 
galaxies per unit redshift detected by the survey and 
\begin{equation}
F(x)^2=1+\Omega_R (H_0x/c)^2.
\end{equation}
The value of $x$ is given by 
\begin{equation}
x=\frac{c}{H_0}\int^z_0 \frac{1}{E(z)}
\end{equation}
where
\begin{equation}
E(z) \equiv \sqrt{\Omega_M (1+z)^3 + \Omega_R(1+z)^2 + \Omega_\Lambda}.
\end{equation}
For values of $z<0.5$, the value of $B$ is more strongly dependent on 
$r_0$, $\gamma$ and the galaxy redshift distribution than the 
cosmological model. This is not unexpected as the value of $x$ for 
$z=0.4$ varies by 
less than $10\%$ between $\Omega_M =1$ and $\Omega_M = 0.2$
models of the Universe. We therefore only use a single cosmological model 
with $H_0 = 75 {\rmn kms}^{-1} {\rmn Mpc}^{-1}$, $\Omega_M=0.2$ and
$\Omega_\Lambda=0$. 

There are two approaches to modelling the galaxy 
redshift distribution which are often used in the literature. 
The first approach uses an accurate description of the local 
luminosity function of different 
galaxy types, plus models of galaxy evolution and {\it k}-corrections 
for each type, and attempts to model the observed number counts and 
redshift distribution where data is available (i.e. Roche {\it et al.} 1996). 
This model includes the physics of galaxy evolution, however, it contains
large numbers of free parameters and different models can readily 
reproduce the
observed number counts. The second approach, which is applied to our 
single band imaging data, 
assumes a functional form for the redshift distribution and 
uses the observed number counts and redshift surveys to constrain the model
(Baugh \& Efstathiou 1993). 
This approach produces a good model of the redshift distribution but
contains no physics of galaxy evolution and is limited by the depth 
of redshift surveys. 

To model the redshift distribution, we use the model 
of Baugh \& Efstathiou (1993). For a complete sample of galaxies brighter
than magnitude $m$, 
the number of galaxies detected per unit redshift is given by 
\begin{equation}
\frac{dN(z)}{dz} = A(m)
z^2 {\rm exp} \left( - \left[ \frac{z}{1.412z_m(m)} \right]^{3/2} \right)
\end{equation}
where
\begin{equation}
A(m)=\frac{1.062 N(m) \Omega }{z_m(m)},
\end{equation} 
$N(m)$ is the galaxy number counts, $\Omega$ is the survey area 
and $z_m$ is the median redshift of the galaxy sample.
As the value of $B$ depends on the galaxy redshift distribution and not
the galaxy number counts, the most important free parameter is $z_m$
which is a function of band and limiting magnitude.

\begin{figure}
\psfig{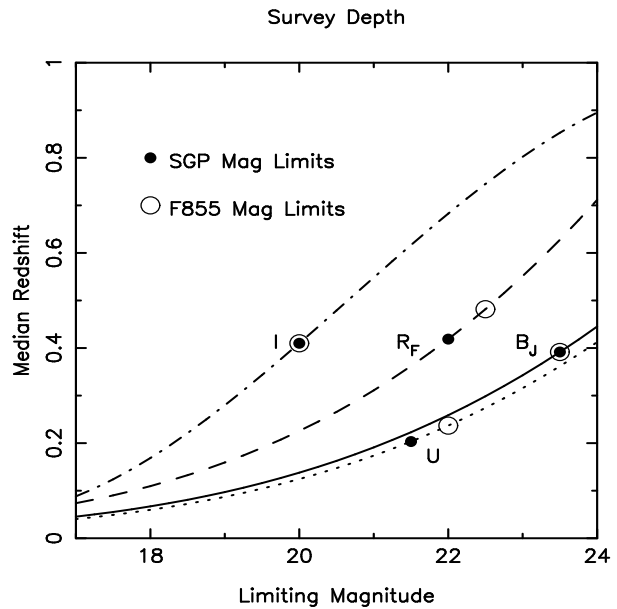}
\caption{A plot of the model median redshift as a function of survey depth 
for $U$, $B_J$, $R_F$ and $I$ bands. The median redshift for the $U$ band 
is derived using a median galaxy colour of $U-B_J\sim 0.3$. 
Magnitude limits for the SGP and F855 are shown with the 
large filled and open circles respectively.}
\label{figure:zmed}
\end{figure}

The median redshift as a function of limiting magnitude 
is obtained from a polynomial fit to median redshifts
derived from galaxy redshift surveys. 
At low redshift this is derived from 
the local galaxy luminosity function while at higher redshifts the 
median redshifts determined directly from the survey data are used. 
Local luminosity functions are assumed to be Schechter functions with 
$\alpha = -1.00$, 
\begin{equation} 
M^*_{B_J}= -19.5-5{\rmn log} h 
\end{equation} 
Loveday {\it et al.} (1992) and 
\begin{equation}
M^*_{R_F}=-20.5 -5{\rmn log} h. 
\end{equation}
Changing the value of $\alpha$ to $-1.3$ 
reduces the estimate of the median redshift by $\sim 15\%$ 
resulting in a similar decrease of the estimate of $r_0$. 
Galaxy $k$-corrections are approximated by $k_B(z)=2z$ and 
$k_R(z)=0.5z$. 
At $z \ga 0.2$, median redshifts as
a function of survey depth are derived from the redshift surveys of 
Colless {\it et al.} (1990), 
Glazebrook {\it et al.} (1995), Lin {\it et al.} (1999) and 
Munn {\it et al.} (1997). $I$ band median redshifts are the same as
Postman {\it et al.} (1998) which were derived from the CFHT redshift
survey (Lilly {\it et al.} 1995). To determine the $U$ band 
median redshift, we use $U-B_J\sim 0.3$, 
the median colour of $B_J<21$ galaxies in the F855 field. The functions
for median redshift as a function of limiting magnitude are shown in 
Figure~\ref{figure:zmed}.  

\section{The Angular Correlation Function at Large Angular Scales}

The $5^{\circ} \times 5^{\circ}$ field-of-view of each field
allows the measurement of the faint galaxy correlation function 
at large angular scales. At $B_J\sim 23.5$ 
and $R_F\sim 22$, the median redshift of the data is $z\sim 0.4$ and 
a break in the spatial correlation function at $\sim 15 h^{-1} {\rmn{Mpc}}$ 
in comoving coordinates (Maddox, Efstathiou \& Sutherland 1996) 
corresponds to an angular scale of $\sim 1 ^{\circ}$. 

It is possible to measure the angular correlation for each field 
to $\sim 5^\circ$ but at large angular scales the 
estimate of the correlation function will be dominated by individual 
structures. To determine the range of angular scales where the correlation 
function is representative, we compare the estimates of the 
angular correlation function for the SGP and F855 fields. This provides
a more reliable estimate than subsamples of the data as each subsample 
would have smaller field-of-view than the original data, resulting 
in overestimates of the errors at large angular scales. 

\begin{figure}
\psfig{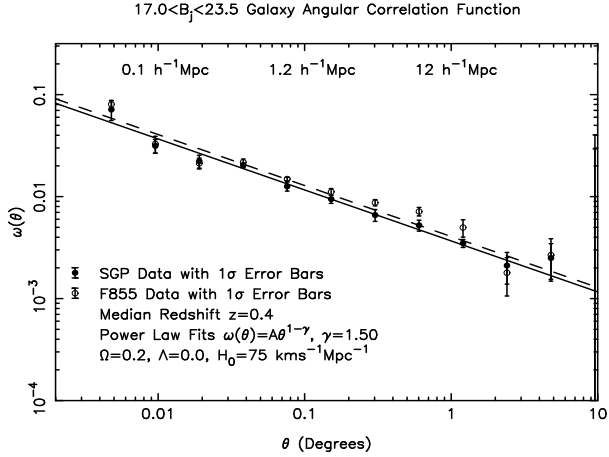}
\caption{The $B_J<23.5$ angular correlation function for
the SGP and F855 fields. Power law fits to the SGP and F855 
data are shown with solid and dashed lines respectively.  
The data in each field is fitted well by 
a power law and most data points are within $2\sigma$ of each other. 
There is no evidence of a break in the correlation function on 
any of the scales measured though $\omega(\theta)$ at $>1^\circ$ is only 
$\sim 2 \sigma$ more than 0. It is also possible both correlation
functions could be biased by large structures on scales $>1^\circ$.}
\label{figure:Jlang}
\end{figure}

\begin{figure}
\psfig{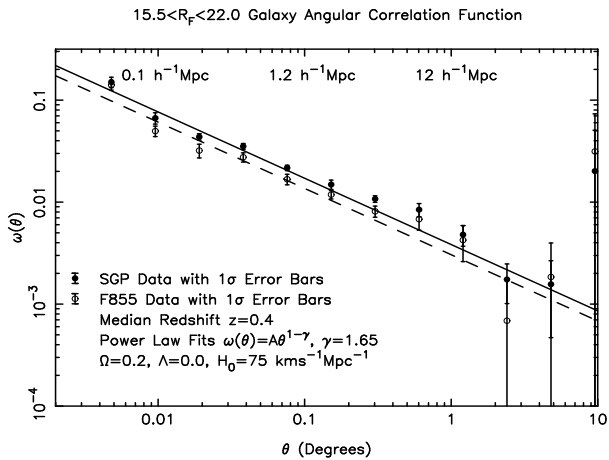}
 \caption{The $R_F<22$ angular correlation function for
the SGP and F855 fields. 
Power law fits to the SGP and F855 
data are shown with solid and dashed lines respectively.  
The data for each field is fitted well by a power law and
there is no evidence of a break in the correlation function on 
scales less than $\sim 2^\circ$.
However, unlike the $B_J<23.5$ data, 
there is an offset between the data in the two fields.}
\label{figure:Rlang}
\end{figure}

The $B_J<23.5$ and $R_F<22$ correlation functions for each field are 
shown in Figures~\ref{figure:Jlang} and~\ref{figure:Rlang}. As the 
integral constraint depends of $\gamma$, the data has been fitted
with power laws with fixed $\gamma$ to allow comparison of the 
2 fields estimates of $\omega(\theta)$. For both fields the integral 
constraint is less than the amplitude of the correlation function at 
$1^\circ$. A power law matches the data well and there is 
no evidence of a break from a power law on all angular scales. 
However, on angular scales 
$\ga 2^\circ$, the estimates of $\omega(\theta)$ for both fields 
are within $2\sigma$ of $0$. It is therefore possible that the break 
in the correlation function is present at $15 h^{-1} {\rmn Mpc}$
in comoving coordinates but can not be detected with this dataset. 

\section{The Correlation Function as a Function of Limiting Magnitude}

The large fields-of-view used for this study reduce the errors associated
with large structures along the line-of-sight. Also, the 
input catalogue of up to $2.5\times 10^5$ galaxies results in small 
random errors in estimates of $\omega (\theta)$. Previous measures
of the correlation function in $U$, $B_J$ and $R_F$ have been restricted to 
fields-of-view less than $\sim 2^\circ \times 2^\circ$ 
(Infante \& Pritchet 1995). At the present time, 
only the $I$ band survey by Postman {\it et al.} (1998) has a 
comparable field-of-view with greater depth than this work. 

Previous measurements of $\gamma$ indicate that it may vary with 
band and survey depth (Infante \& Pritchet 1995, Postman {\it et al.} 1998) 
though most values in the literature are between $1.6$ and $1.8$.
The consistent data reduction method used for this work should
allow the accurate comparison of $\gamma$ as a function of band
and survey depth. 

\begin{figure}
\psfig{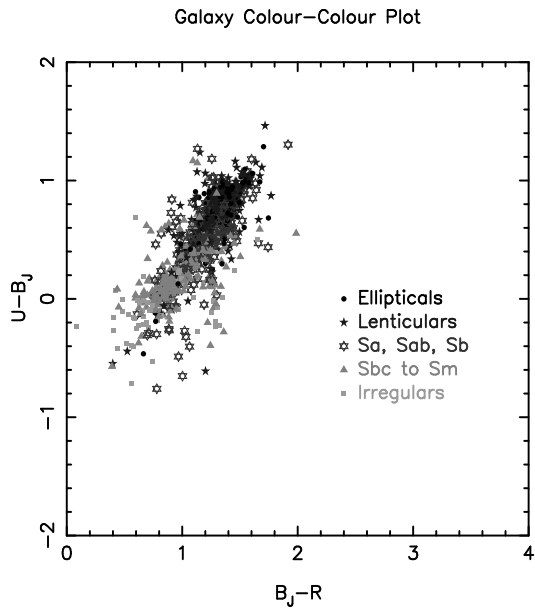}
\caption{The colours of RC3 catalogue (de Vaucouleurs {\it et al.} 1991) 
galaxies by morphological type. Photometry is from 
Prugniel \& Heraudeau (1998) while morphological classification of galaxies
is from de Vaucouleurs {\it et al.} (1991). 
The correlation between morphology and colour can be clearly seen with 
late type galaxies generally having bluer colours than early types.} 
\label{figure:rc3}
\end{figure}

Estimates of the value of $\omega(1^\prime)$ and $\gamma$ derived 
from power law fits to the data are shown in Table~\ref{table:wp}. 
The amplitude of the correlation function is determined at 
$1^\prime$ rather than $1^\circ$ as the amplitude and $\gamma$ are not 
independent and estimates of $\omega(1^\circ)$ depend strongly on $\gamma$. 
The values of $\gamma$ show a weak trend towards smaller
values with magnitude and bluer survey bands. As there is a correlation 
between galaxy colour and morphology (see Figure~\ref{figure:rc3}) and
late type galaxies have shallower values of $\gamma$ than early type 
galaxies, this trend is not unexpected.

\begin{table*}
\caption{Measured  parameters for $\omega(\theta)$ determined 
from power law fits to data between $10^{\prime \prime}$ and $0.3^\circ$.}
\label{table:wp}
\begin{tabular}{lrcrrcr}
Field & \multicolumn{3}{c}{SGP} &  \multicolumn{3}{c}{F855} \\
\\
Magnitude Range & $N_{gal}$ & $\gamma$ 	 & $\omega(1^\prime) \times 10^3$  	& $N_{gal}$ & $\gamma$ 	 & $\omega(1^\prime) \times 10^3$  \\
\\
$18.0 \leq U \leq 20.0$ & 4309  & $1.39\pm 0.18$ & $418\pm 151$			& 4913 & $1.53\pm 0.27$ & $239\pm 52$	\\
$18.0 \leq U \leq 21.0$ & 16379 & $1.65\pm 0.10$ & $196\pm 16$			& 16695 & $1.84\pm 0.10$ & $125\pm 16$ \\
$18.0 \leq U \leq 22.0$ & 80502 & $1.47\pm 0.10$ & $63\pm 4$			& 82965 & $1.47\pm 0.15$ & $37\pm 5$ \\
\\
$18.0 \leq B_J \leq 20.0$ & 5267  & $1.58\pm 0.15$ & $530 \pm 85$ 		& 5486 & $1.76\pm 0.17$ & $335 \pm 62$ \\
$18.0 \leq B_J \leq 21.0$ & 15372 & $1.60\pm 0.12$ & $280 \pm 32$ 		& 15149 & $1.72\pm 0.12$ & $196 \pm 27$ \\
$18.0 \leq B_J \leq 22.0$ & 45519 & $1.53\pm 0.07$ & $122 \pm 7$ 		& 43743 & $1.66\pm 0.11$ & $90 \pm 8$ \\
$18.0 \leq B_J \leq 22.5$ & 82919 & $1.48\pm 0.09$ & $72 \pm 5$ 		& 76149 & $1.60\pm 0.07$ & $56 \pm 4$ \\
$18.0 \leq B_J \leq 23.0$ & 144577 & $1.53\pm 0.07$ & $45 \pm 2$ 		& 123325 & $1.47\pm 0.08$ & $38 \pm 2$ \\
$18.0 \leq B_J \leq 23.5$ & 230024 & $1.46\pm 0.07$ & $28 \pm 1$ 		& 184717 & $1.49\pm 0.05$ & $31 \pm 2$ \\
\\
$16.5 \leq R_F \leq 18.5$ & 4330   & $1.68\pm 0.10$ & $586 \pm 80$ 		& 4294 & $1.82\pm 0.17$ & $397\pm 82$ \\
$16.5 \leq R_F \leq 19.5$ & 13752  & $1.72\pm 0.07$ & $346 \pm 50$ 		& 12841 & $1.81\pm 0.13$ & $232\pm 32$ \\
$16.5 \leq R_F \leq 20.5$ & 39748  & $1.70\pm 0.03$ & $179 \pm 9$ 		& 37069 & $1.80\pm 0.08$ & $130\pm 12$ \\		
$16.5 \leq R_F \leq 21.5$ & 112919 & $1.58\pm 0.05$ & $72 \pm 3$ 		& 102373 & $1.70\pm 0.07$ & $61\pm 4$ \\		
$16.5 \leq R_F \leq 22.0$ & 174385 & $1.57\pm 0.05$ & $54 \pm 2$ 		& 173295 & $1.60\pm 0.05$ & $43\pm 2$ \\	
$16.5 \leq R_F \leq 22.5$ &	-  &		-   &		-		& 234957 & $1.65\pm 0.07$ & $32\pm 2$ \\ 
\\
$16.5 \leq I \leq 18.5$ & 9812  & $1.73 \pm 0.07$ & $322\pm 26$ 		& 8913 & $1.82\pm 0.15$ & $254\pm 41$ \\
$16.5 \leq I \leq 19.5$ & 32178 & $1.67 \pm 0.06$ & $156\pm 11$ 		& 28365 & $1.79\pm 0.12$ & $115\pm 14$ \\
$16.5 \leq I \leq 20.0$ & 54737 & $1.71 \pm 0.05$ & $108 \pm 8$ 		& 48743 & $1.80\pm 0.08$ & $93\pm 8$ \\
\\
\end{tabular}
\end{table*}

While the value of $\gamma$ does not differ by more 
then $\sim 2 \sigma$ between the two fields, the amplitude of the 
correlation function varies by $\sim 100\%$ at bright magnitudes. While
variations of amplitude could be caused by zero point errors, the error 
required is approximately $0.4$ magnitudes in all bands at bright 
magnitudes with it decreasing to $\sim 0$ at $B_J\sim 23$. 
This is inconsistent with the photometric calibration, 
the galaxy number counts in Figure~\ref{figure:counts} and 
the colour-colour diagrams. A systematic error could be 
present in the data but it seems unlikely that it 
could effect $\omega(1^\prime)$ without causing large variations of $\gamma$. 

If the values 
of $\omega(1^\prime)$ do not have significant systematic errors, a 
possible cause of the variations is that the two fields measure
different populations of galaxies with different clustering properties. 
The SGP, which has stronger clustering than F855, contains 
``sheets'' perpendicular to the line-of-sight
(Broadhurst {\it et al.} 1990) and smaller structures including 
22 Abell clusters (Abell, Corwin \& Olowin 1989). Of these clusters, 
11 appear to be associated with a structure at $z\sim 0.11$ identified by 
Broadhurst {\it et al.} (1990). $M_{B_J}\sim -19.5$ ($M^*$) galaxies
at $z\sim 0.11$ have an apparent magnitude of $B_J \sim 18$ and this
population of galaxies is too small to significantly bias estimates of the 
$B_J>20$ correlation function. However, it is possible that the 
cluster population of $M_{B_J}\ga -16$ galaxies could bias 
estimates of the correlation function if they are a significant fraction
of the observed galaxy number counts.

To test if the $z=0.11$ clusters do significantly effect the 
correlation function, the correlation function has been determined with 
$2^\circ \times 2^ \circ$ 
($\sim 10 h^{-1} {\rmn Mpc} \times 10 h^{-1} {\rmn Mpc}$)
regions surrounding the clusters 
removed from the data. As the clusters are not uniformly distributed across
the field-of-view, the size of the catalogue is reduced by $\sim 60\%$. 
Table~\ref{table:noclust} lists
the amplitude of the $B_J$ and $R_F$ band correlation functions 
for the SGP field without the clusters and F855 for comparison. 
The amplitude of the correlation function has decreased significantly compared 
with the original estimates for the SGP. While most noticeable at bright
magnitudes, the effect is also significant at fainter magnitudes where 
the contribution from nearby clusters might be expected to be small. 
This is consistent with $M_{B_J}\ga -16$ galaxies within the clusters 
significantly effecting estimates of the faint galaxy correlation function. 
However, details of the relationship between clusters and the observed 
clustering of faint galaxies will be explored in a later paper.  

\begin{table*}
\caption{Measured  parameters for $\omega(\theta)$ determined 
from power law fits to data between $10^{\prime \prime}$ and $0.3^\circ$.
Clusters at $z \sim 0.11$ have been removed from the SGP sample to reduce the
effect dwarf galaxies may have on the estimate of the correlation function.}
\label{table:noclust}
\begin{tabular}{lrcrrcr}
Field & \multicolumn{3}{c}{SGP (no $z=0.11$ clusters)} &  \multicolumn{3}{c}{F855} \\
\\
Magnitude Range & $N_{gal}$ & $\gamma$ 	 & $\omega(1^\prime) \times 10^3$  	& $N_{gal}$ & $\gamma$ 	 & $\omega(1^\prime) \times 10^3$  \\
\\
$18.0 \leq B_J \leq 20.0$ & 1742  & $1.95\pm 0.35$ & $530 \pm 85$ 		& 5486 & $1.76\pm 0.17$ & $335 \pm 62$ \\
$18.0 \leq B_J \leq 21.0$ & 5487  & $1.76\pm 0.20$ & $200 \pm 44$ 		& 15149 & $1.72\pm 0.12$ & $196 \pm 27$ \\
$18.0 \leq B_J \leq 22.0$ & 16679 & $1.72\pm 0.12$ & $108 \pm 11$ 		& 43743 & $1.66\pm 0.11$ & $90 \pm 8$ \\
$18.0 \leq B_J \leq 22.5$ & 30886 & $1.61\pm 0.17$ & $60 \pm 8$ 		& 76149 & $1.60\pm 0.07$ & $56 \pm 4$ \\
$18.0 \leq B_J \leq 23.0$ & 54549 & $1.54\pm 0.15$ & $36 \pm 4$ 		& 123325 & $1.47\pm 0.08$ & $38 \pm 2$ \\
$18.0 \leq B_J \leq 23.5$ & 87453 & $1.56\pm 0.16$ & $22 \pm 3$ 		& 184717 & $1.49\pm 0.05$ & $31 \pm 2$ \\
\\
$16.5 \leq R_F \leq 18.5$ & 1460   & $2.16\pm 0.14$ & $339 \pm 170$ 		& 4294 & $1.82\pm 0.17$ & $397\pm 82$ \\
$16.5 \leq R_F \leq 19.5$ & 4974   & $1.84\pm 0.18$ & $223 \pm 30$ 		& 12841 & $1.81\pm 0.13$ & $232\pm 32$ \\
$16.5 \leq R_F \leq 20.5$ & 14481  & $1.76\pm 0.13$ & $144 \pm 12$ 		& 37069 & $1.80\pm 0.08$ & $130\pm 12$ \\		
$16.5 \leq R_F \leq 21.5$ & 41767  & $1.67\pm 0.11$ & $65 \pm 6$ 		& 102373 & $1.70\pm 0.07$ & $61\pm 4$ \\		
$16.5 \leq R_F \leq 22.0$ & 65168  & $1.71\pm 0.10$ & $49 \pm 2$ 		& 173295 & $1.60\pm 0.05$ & $43\pm 2$ \\	
\\
\end{tabular}
\end{table*}

For the following discussion of the amplitude as a function of limiting
magnitude, the estimates of the SGP correlation function
including the $z=0.11$ clusters are used. While it is probable
relatively nearby clusters do effect 
estimates of the correlation function, there is no clear justification
for excluding them.
Also, excluding the regions surrounding the clusters
significantly reduces the number of galaxy pairs used to determine the
correlation function, significantly increasing random errors. 

\begin{figure}
\psfig{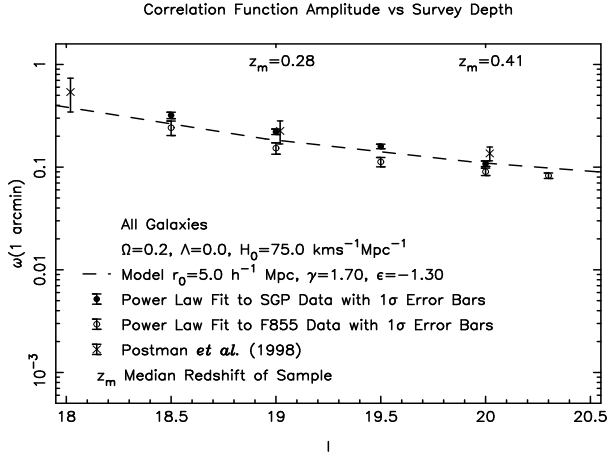}
\caption{The $I$ band correlation function amplitude. The measurements
of the amplitude are consistent with the data of Postman {\it et al.} (1998)
which is shown with crosses. The data from Postman {\it et al.} (1998)
has been corrected for the assumption that $\gamma=1.7$.}
\label{figure:Iampmag}
\end{figure}

\begin{figure}
\psfig{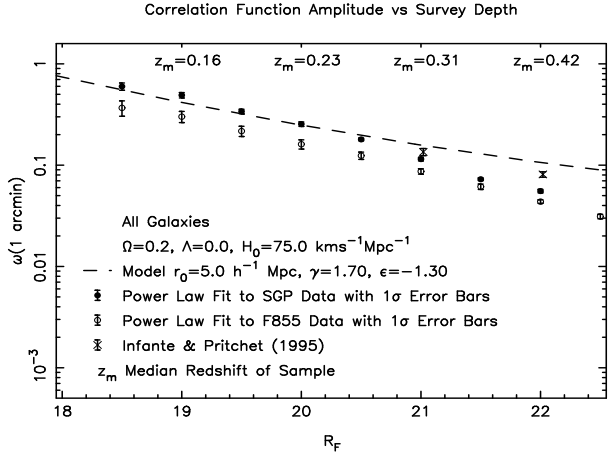}
\caption{The $R_F$ band correlation function amplitude. While the redshift
range is similar to the $I$ band data, the galaxy clustering is significantly
weaker.}
\label{figure:Rampmag}
\end{figure}

\begin{figure}
\psfig{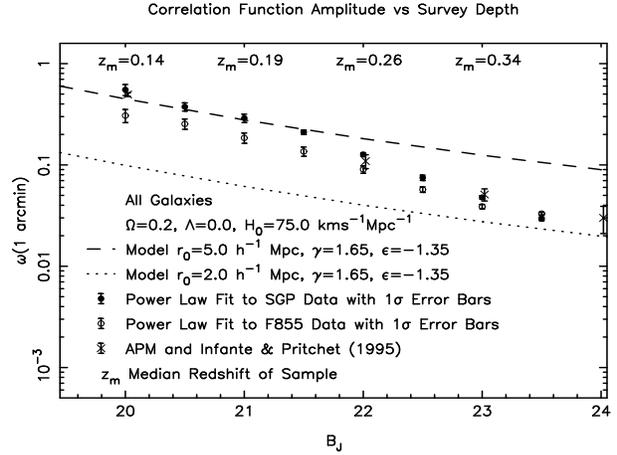}
\caption{The $B_J$ band correlation function amplitude. As with 
the $R_F$ band data, the amplitude of the correlation function is 
significantly weaker than the $I$ band correlation function. At
magnitudes fainter than $B_J\sim 22$, the correlation function rapidly
decreases with limiting magnitude. There is also a decrease in the 
discrepancy between the SGP and F855 measurements with increasing
limiting magnitude.}
\label{figure:Jampmag}
\end{figure}

\begin{figure}
\psfig{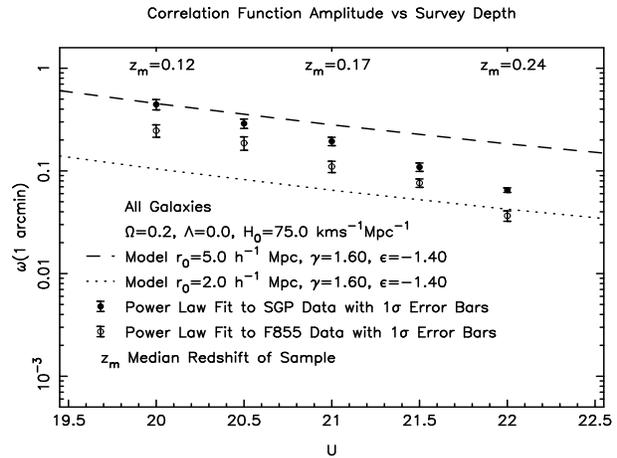}
\caption{The $U$ band correlation function amplitude. The amplitude of 
the correlation function is consistently weaker than the fit to the $I$
band data shown with the dashed line.}
\label{figure:Uampmag}
\end{figure}

Figures~\ref{figure:Iampmag}~to~\ref{figure:Uampmag}
plot the $I$, $R_F$, $B_J$ and $U$ band angular correlation function 
amplitudes as a function of limiting magnitude.
The amplitude of the correlation function has been determined 
with fixed values of $\gamma$ to reduce the dependence of the amplitude
on $\gamma$. 
As the values of the angular correlation function differ significantly
between each field, no attempt has been made to fit the data. Instead, 
a model has been plotted with clustering fixed in comoving coordinates
and $r_0 = 5 h^{-1} {\rm Mpc}$. The value of $r_0$ is similar to 
values derived by Maddox, Efstathiou \& Sutherland (1996) 
and Postman {\it et al.} (1998). 

Figure~\ref{figure:Iampmag} shows the measured amplitude of the
$I$ band correlation function for the SGP, F855 and 
Postman {\it et al.} (1998). The measured clustering in 
the SGP is in good agreement with Postman {\it et al.} (1998), while
F855 measures slightly weaker clustering. The model of the clustering
is also a reasonable estimate of the observed clustering across the 
magnitude range observed. 
Figures~\ref{figure:Rampmag} to~\ref{figure:Uampmag} show
the $R_F$, $B_J$ and $U$ band clustering to be significantly weaker than the
$I$ band clustering for galaxies with a similar range of redshifts. 
Comparison with the results of Infante \& Pritchet (1995) show
the SGP has similar clustering in $B_J$ and weaker clustering in 
$R_F$. Infante \& Pritchet (1995) may 
measure stronger clustering than F855 as their field is located
in the NGP which has similar large scale structures as the SGP
(Broadhurst {\it et al.} 1990). 

While the $r_0=5.0 h^{-1} {\rmn Mpc}$
model is a good fit to the SGP $R_F<20$ and $B_J<21$ data, 
at fainter magnitudes a rapid decline in the amplitude of the correlation
function is observed. The rapid decline of the faint correlation function
has been observed previously by  Efstathiou {\it et al.} (1991), 
Infante \& Pritchet (1995) and Roche {\it et al.} (1996). 
Obviously, the $R_F$ and $B_J$ band data are
sampling a different population of galaxies to the $I$ band sample. 
To be consistent with galaxy redshift surveys, the $B_J\sim 23$ galaxies
must be dominated by a population of weakly 
($r_0\sim 2.0 h^{-1} {\rmn Mpc}$) clustered galaxies at 
$z \sim 0.4$ (Efstathiou {\it et al.} 1991, Efstathiou 1995). 
  
For a no-evolution model for the galaxy population, the blue bands 
will sample galaxies with bluer colours due to the large {\it k}-corrections
of early-type galaxies (Coleman, Wu \& Weedman 1980). 
As shown in Figure~\ref{figure:rc3}, the local population of blue 
($U-B_J \la 0$) galaxies is dominated by late-type galaxies. 
Measurements of the galaxy clustering in the local universe with morphology-selected 
catalogues show the clustering of late-type galaxies is considerably
weaker than early-type galaxies (Davis \& Geller 1976, 
Loveday {\it et al.} 1995). If the colours and 
clustering of $z \sim 0.4$ late and early type galaxies are similar
to $z \sim 0$ galaxies, the decrease in the amplitude of the $B_J$ 
angular correlation function could be a selection effect. While 
it is impossible to determine the morphologies of the $B_J>21$ galaxies 
with this catalogue, it should be possible to use colour selection to 
select early and late type galaxies over a range of redshifts.  

\section{The Correlation Function of Colour Selected Galaxies}

\begin{figure}
\psfig{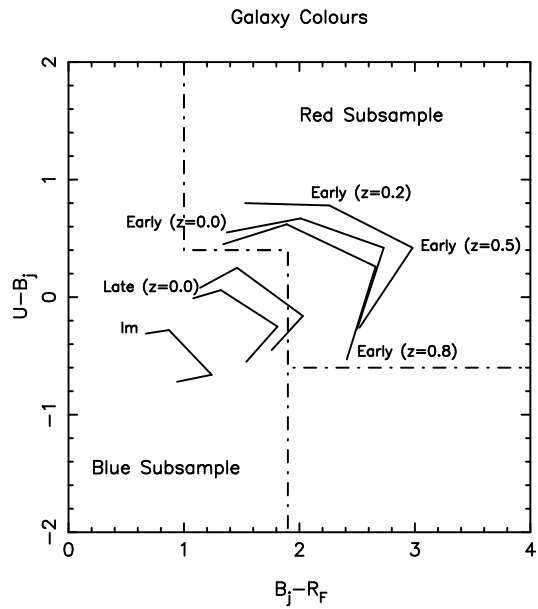}
\caption{The colour selection criteria for the red and blue subsamples. 
Colours of early, late and irregular type galaxies at $z=0$, $z=0.2$, 
$z=0.5$ and $z=0.8$ from Fukugita, Shimasaku \& Ichikawa (1995) are shown. 
The colour selection criteria for each subsample are shown with the 
dot-dash line. The blue subsample consists of galaxies below and to the
left of the line while the red subsample consists of galaxies above 
and to the right of the line. The blue subsample is dominated by 
late and irregular type galaxies while the red subsample
selects early type galaxies.}
\label{figure:fukugita}
\end{figure}

The obvious colour selection criteria for galaxies is a single colour cut 
in the deepest bands available ($B_J$ and $R_F$). However, as shown in 
Figure~\ref{figure:fukugita}, the colour of individual galaxy types 
varies with redshift. Even with a single colour cut, 
the blue subsample will generally select later type galaxies 
than the red subsample and blue subsamples generally show weak clustering 
(Infante \& Pritchet 1995, Roche {\it et al.} 1996). A $B_J-R_F$ 
cut that varies with magnitude may select the same
population over a range of redshifts but the selection criteria would depend 
on the cosmological parameters used. 
Also, colour selection of a fraction of galaxies 
(i.e. the reddest $20\%$ of the catalogue) may not be effective 
due to the changing morphological mix of galaxies with magnitude. 

\begin{figure*}
\psfig{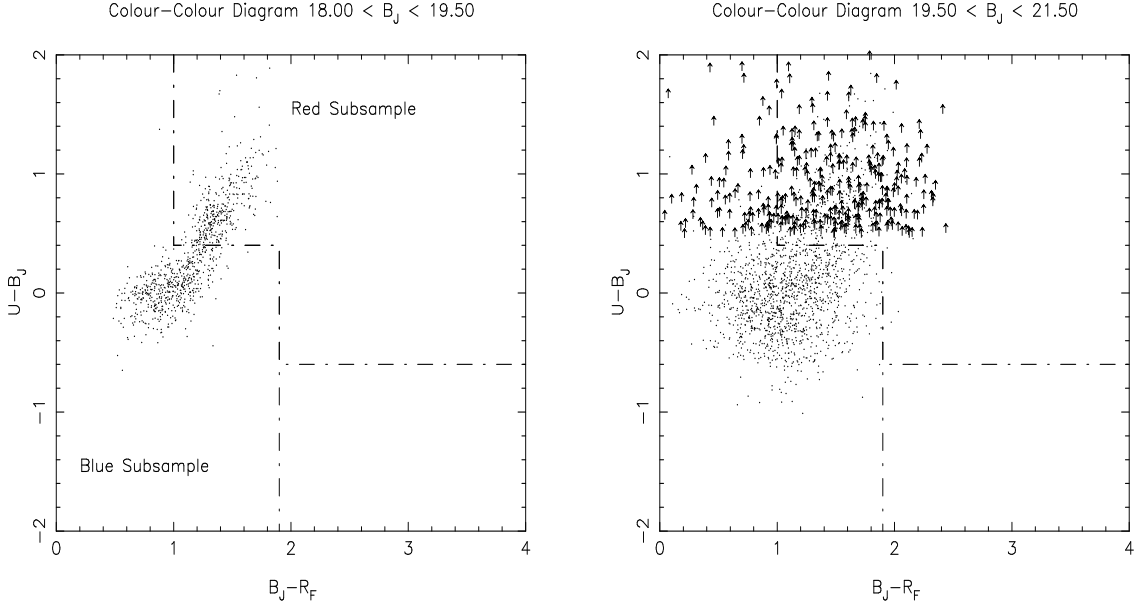}
\caption{Galaxy colour-colour diagrams for the SGP shown with the selection criteria 
for the two subsamples.}
\label{figure:SGPcolcol}
\end{figure*}

\begin{figure}
\psfig{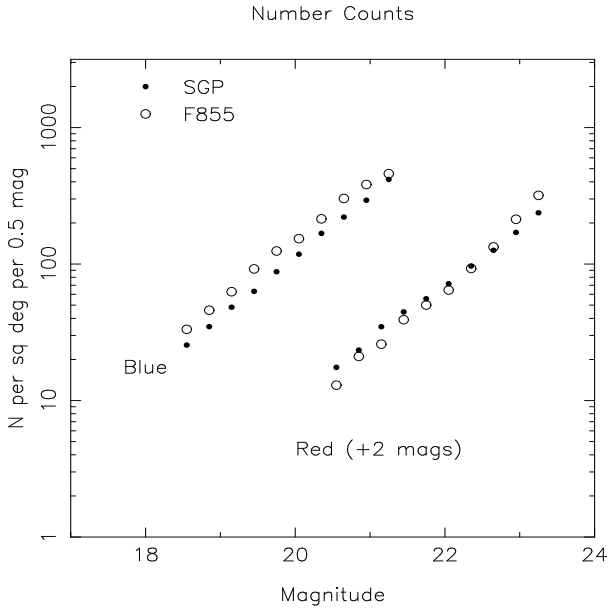}
\caption{A plot of the red and blue 
subsample number counts for the SGP and F855.}
\label{figure:colcounts}
\end{figure}

Selection of galaxies with two or more colour criteria should allow the 
selection of galaxy types over a large range of redshifts without 
strong dependence on cosmological parameters. Figure~\ref{figure:fukugita} 
shows the predicted colours of early, late and irregular type 
$0  \leq z \leq 0.8$ galaxies from Fukugita, Shimasaku \& Ichikawa (1995). 
Colour selection criteria for a red and blue subsample are also shown. 
The samples are limited to $B_J<21.6$ due to the $U=22$ magnitude
limit of the catalogues and the $U-B_J>0.4$ selection criteria. 
As most $B_J<21.6$ galaxies are at redshifts $z\leq 0.3$, most galaxies
in the subsamples are selected with the $U-B_J$ selection criteria. 
Galaxies with $U-B_J$ lower limits and $B_J-R_F$ upper limits have
been included in the subsamples to prevent incompleteness. 
Figure~\ref{figure:SGPcolcol} shows the colours of galaxies in the SGP 
with the subsample selection criteria. The location of the
$B_J<19.5$ galaxy locus is similar to that for RC3 catalogue galaxies in
Figure~\ref{figure:rc3} though at $B_J>19.5$
there is an increasing fraction of very blue galaxies. Galaxy number counts 
for the 2 subsamples are shown in Figure~\ref{figure:colcounts}. While
the SGP does include significantly more clusters than F855 at low redshift,
there is good agreement between the number counts across the magnitude 
range observed. 

\begin{figure}
\psfig{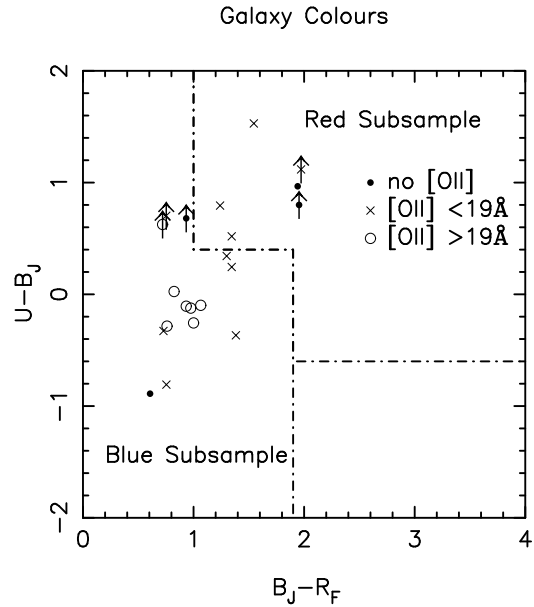}
\caption{A colour-colour diagram of $B_J<21.5$ 
galaxies in the SGP and F855 with 
spectroscopic observations by Colless {\it et al.} (1990). Lower limits in 
$U-B_J$ are shown with arrows. A trend towards
bluer colours with $[{\rmn OII}]~3727 {\rmn \AA}$ emission can 
be observed with strong $[{\rmn OII}]~3727 {\rmn \AA}$ emitters
restricted to the blue subsample.}
\label{figure:colless}
\end{figure}

While the selection criteria are relatively simple, it can be clearly seen 
that the red and blue subsample should select 
early and late type galaxies respectively. The blue subsample should
also contain most of the galaxies with significant star formation rates 
while the red subsample should contain more passive galaxies. 
As shown in Figure~\ref{figure:colless}, 
comparison with the spectroscopic catalogue of Colless 
{\it et al.} (1990) shows that most of the 
$[{\rmn OII}]~3727 {\rmn \AA}$ emitters detected 
in the SGP and F855 are included in the blue subsample.
To measure the clustering of starforming galaxies Cole {\it et al.} (1994) selected galaxies with 
$[{\rmn OII}]~3727 {\rmn \AA}$ equivalent widths greater than 
$19{\rmn \AA}$; all galaxies matching this
selection criteria would be included in the blue subsample for both fields.  
Interestingly, the galaxies without $[{\rmn OII}]~3727 {\rmn \AA}$ emission do not show such 
an obvious trend with similar numbers in both subsamples. 
However, the small number of galaxies without $[{\rmn OII}]~3727 {\rmn \AA}$ emission results in 
these galaxies comprising less than $15\%$ of the total of blue galaxies.

\begin{figure}
\psfig{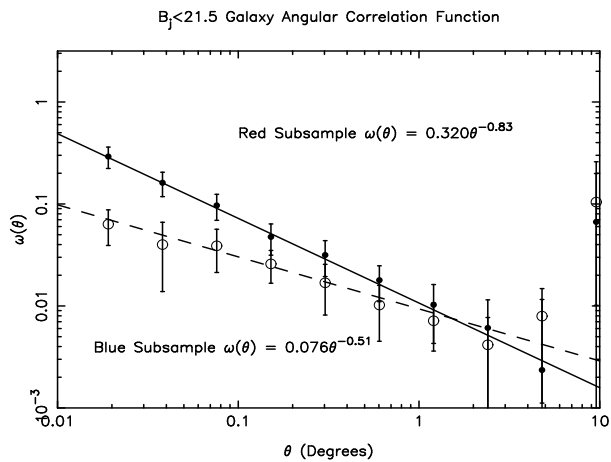}
\caption{The $B_J<21.5$ angular correlation function for the red and 
blue subsamples of F855. The red subsample shows stronger clustering and 
a higher value of $\gamma$, consistent with the strong clustering of 
elliptical and early type galaxies in the local universe.}
\label{figure:redblue}
\end{figure}

The $B_J<21.5$ angular correlation functions of the blue and 
red subsamples of F855 are shown in Figure~\ref{figure:redblue}. 
Visual inspection shows the significant difference in clustering strength
and the value of $\gamma$ for the two subsamples with the red subsample 
being strongly clustered and having a higher value of $\gamma$. 
This trend is consistent with measurements of the angular correlation 
functions of morphology selected catalogues where $\gamma \sim 1.8$ 
for early-type galaxies and $\gamma \sim 1.5$ for late-type galaxies 
(Loveday {\it et al.} 1995).

\begin{table*}
\caption{Measured  parameters for $\omega(\theta)$ for the 
red and blue subsamples determined 
from power law fits to data between $10^{\prime \prime}$ and $0.3^\circ$.}
\label{table:ct}
\begin{tabular}{llrcrrcr}
\multicolumn{2}{c}{Field} & \multicolumn{3}{c}{SGP} &  \multicolumn{3}{c}{F855} \\
\\
Sample & Magnitude Range & $N_{gal}$ & $\gamma$ 	 & $\omega(1^\prime) \times 10^3$  & $N_{gal}$ & $\gamma$ & $\omega(1^\prime) \times 10^3$  \\
\\
Blue & $18.0 \leq B_J \leq 20.0$ & 2963  & $1.45\pm 0.51$ & $212\pm 3200$		& 3354 & $1.41\pm 0.58$ & $170\pm 2710$	\\
Blue & $18.0 \leq B_J \leq 21.0$ & 9079  & $1.49\pm 0.18$ & $138\pm 27$			& 9228 & $1.61\pm 0.38$ & $110\pm 39$ \\
Blue & $18.0 \leq B_J \leq 21.5$ & 15579 & $1.57\pm 0.20$ & $86\pm 19$			& 14329 & $1.51\pm 0.32$ & $75\pm 32$ \\
\\
Red & $18.0 \leq B_J \leq 20.0$ & 1993 & $1.65\pm 0.22$ & $1339\pm 278$			& 1851 & $2.09\pm 0.17$ & $954\pm 210$	\\
Red & $18.0 \leq B_J \leq 21.0$ & 5552 & $1.69\pm 0.12$ & $782\pm 73$			& 5424 & $1.89\pm 0.13$ & $480\pm 82$   \\
Red & $18.0 \leq B_J \leq 21.5$ & 9317 & $1.72\pm 0.08$ & $616\pm 51$			& 9922 & $1.83\pm 0.14$ & $319\pm 44$  \\
\\
\end{tabular}
\end{table*}

Estimates of $\gamma$ and $\omega(1^\prime)$ as a function of 
limiting magnitude are listed in Table~\ref{table:ct}. The $z\sim 0.11$
clusters in the SGP have been retained in the data as there is no 
obvious justification for rejecting them from the sample. 
While $\gamma$
differs significantly between the two subsamples, the values of $\gamma$ are 
within $1\sigma$ of being constant as a function of limiting magnitude.
The values of $\omega(1^\prime)$ for the blue subsamples are remarkably
similar for the SGP and F855. However, the red subsamples show 
large differences with the amplitude of the clustering in the 
SGP being $\sim 100\%$ larger than F855. 
 
To model the spatial correlation function, a model of the redshift
distribution for the red and blue subsamples is required. Luminosity
functions for galaxies selected by morphology are available but
it is not clear if luminosity is more or less strongly correlated with 
colour than morphology. We therefore use the $U-B_J<0.2$ and 
$U-B_J>0.2$ galaxy luminosity functions of Metcalfe {\it et al.} (1998).
Approximate $k-$corrections of $k(z)=3z$ and $k(z)=z$ are used for 
the red and blue subsample respectively. 
We use the Metcalfe {\it et al.} (1998) luminosity functions 
rather than the redshift distribution model as the steep slope of 
faint end of the luminosity functions results in skewed redshift 
distributions. 

\begin{figure}
\psfig{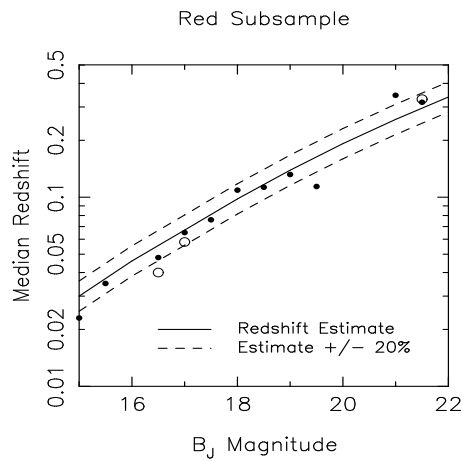}
\caption{The median redshift as a function of magnitude for the red 
subsample. Data points from $B_J$ band magnitude 
limited surveys are shown with circles while data using 
NED redshifts is shown with dots. The median
redshift estimate and data points have been determined using
0.5 magnitude wide bins. As the estimates of the correlation function
use wider magnitude bins, the estimates of the median redshift in 
this plot are higher than in  Figure~\ref{figure:redampmag}.}
\label{figure:redz}
\end{figure}

\begin{figure}
\psfig{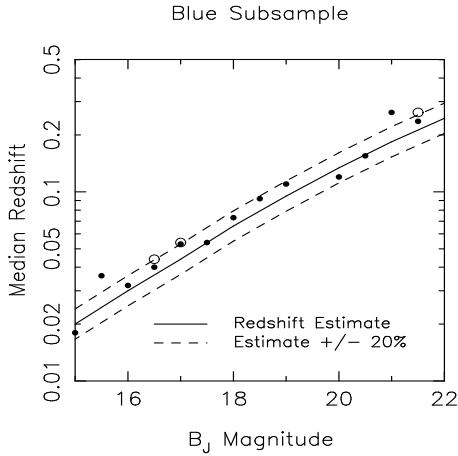}
\caption{The median redshift as a function of magnitude for the blue 
subsample. Data points from $B_J$ band 
magnitude limited surveys are shown with 
circles while data using NED redshifts is shown with dots 
is shown with dots. The median
redshift estimate and data points have been determined using
0.5 magnitude wide bins. As the estimates of the correlation function
use wider magnitude bins, the estimates of the median redshift in 
this plot are higher than in  Figure~\ref{figure:blueampmag}.}
\label{figure:bluez}
\end{figure}

Plots of the estimated median redshift for the red and blue luminosity
functions are shown in Figures~\ref{figure:redz} and~\ref{figure:bluez}. 
The SGP and/or F855 overlap the $B_J$ magnitude 
limited redshift surveys of Colless {\it et al.} (1990) and 
Ratcliffe {\it et al.} (1998) which have been used to measure
the median redshifts of the red and blue subsamples as a function of 
magnitude. In addition, $\sim 700$ galaxy redshifts are available
from the NASA/IPAC Extragalactic Database (NED) and these have 
been used to show
the {\it approximate} median redshift as a function of magnitude. 
For both subsamples, the model is a reasonable estimate of the redshift 
with few data points being more than $20\%$ from the model redshift 
estimate. 

\begin{figure}
\psfig{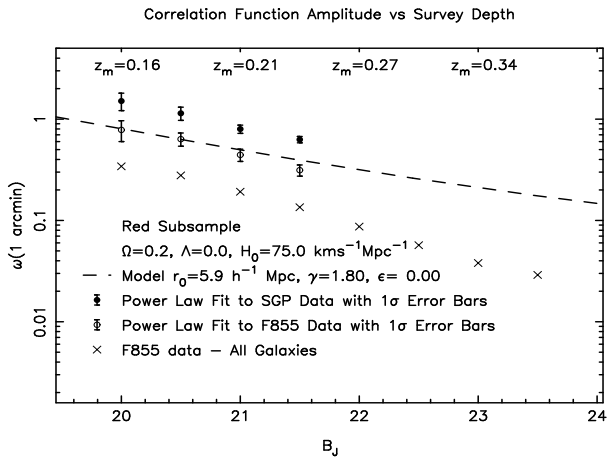}
\caption{The amplitude of the clustering of the red subsamples 
as a function of magnitude. The value of $\gamma$ has been fixed to $1.8$. 
The clustering strength in the two fields differs by $\sim 100\%$ 
at all magnitudes. Both fields show strong clustering, consistent
with the red subsamples being dominated by early type galaxies.}
\label{figure:redampmag}
\end{figure}

\begin{figure}
\psfig{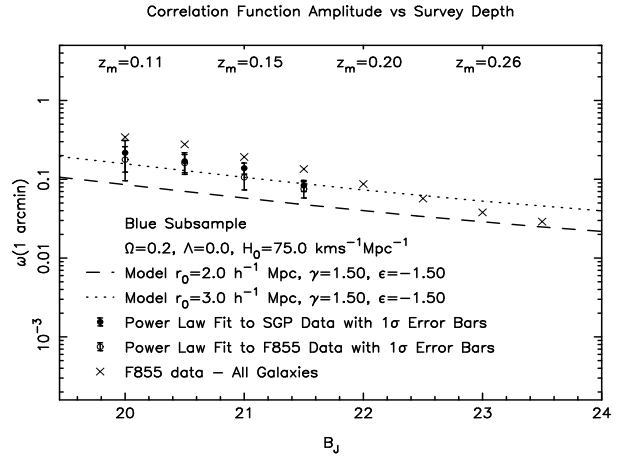}
\caption{The amplitude of the clustering for the blue subsamples
as a function of magnitude. The value of $\gamma$ has been fixed to $1.5$.
The blue galaxies are extremely weakly clustered with a similar
spatial correlation function to faint $B_J$ galaxies. 
As the $B_J \sim 23$ number counts are dominated by 
blue galaxies (Guhathakurta, Tyson \& Majewski 1990), this
is strong evidence for star forming galaxies being weakly clustered
from $z \sim 0.4$ until the present epoch.}
\label{figure:blueampmag}
\end{figure}

Figure~\ref{figure:redampmag} shows the amplitude of the SGP and F855
red subsamples as a function of limiting magnitude
with $\gamma$ fixed at $1.8$.  A model of the early type galaxy 
correlation function with $r_0=5.9 h^{-1} {\rmn Mpc}$  
(Loveday {\it et al.} 1995) and clustering fixed in physical coordinates
($\epsilon=0$)
is also shown. Clustering fixed in physical coordinates would be
applicable to galaxies in gravitationally bound clusters.  
The model, shown with the dashed line, does not fit either set of 
data but this is not unexpected as the SGP and F855 fields 
exhibit significantly different clustering amplitudes. 
However, the clustering in both fields is consistently stronger
than the clustering of all $B_J$ galaxies in F855. This is consistent
with the red subsample being dominated by strongly clustered 
early type galaxies. 

Figure~\ref{figure:blueampmag} plots the amplitude of the blue subsample
correlation function derived from power law fits with $\gamma$ fixed to
$1.5$. Unlike the red subsample, the amplitudes derived from the 
SGP and F855 data agree at all magnitudes. If dwarf galaxies 
are increasing the amplitude of the SGP correlation function, this 
implies that they are red galaxies such as dwarf ellipticals 
within the clusters. A model with extremely 
weak clustering ($r_0 = 2 h^{-1} {\rmn Mpc}$) fixed in 
comoving coordinates is also shown in 
Figure~\ref{figure:blueampmag}. The amplitude of the clustering is similar
to that estimated for faint galaxies by Efstathiou (1995)
and $z\sim 0.5$ galaxies by Le F\'evre {\it et al.} (1996). 
The model slightly underestimates the strength of the clustering 
and a model with stronger clustering, $r_0 = 3 h^{-1} {\rmn Mpc}$,
is a better fit to the data. This is significantly weaker 
than $r_0\sim 4.4 h^{-1} {\rmn Mpc}$, the measured clustering of 
late type galaxies in the local Universe (Loveday {\it et al.} 1995). 
The only large sample of low $z$ galaxies with similar properties 
($r_0\sim 3 h^{-1} {\rmn Mpc}$, $\gamma\sim 1.8$) 
are galaxies with $[{\rmn OII}]$ or ${\rmn H}\alpha$ emission lines 
with large equivalent widths from the Stromlo-APM survey (Loveday, 
Tresse \& Maddox 1999). However, the Stromlo-APM sample has significantly 
stronger clustering than the blue subsample or faint blue galaxies on 
scales $\la 2  h^{-1} {\rmn Mpc}$.

An underestimate of the redshift by $\ga 50\%$ 
could explain the weak clustering at 
$B_J \sim 20$ but this would be difficult to reconcile with the 
redshift data in Figure~\ref{figure:bluez}. Altering the assumed
cosmological parameters can change estimates of spatial correlation function
but the effect is negligible at $z\sim 0$ and is less than $30\%$ at $z=0.2$. 
A plausible explanation is that the clustering of galaxies 
is more strongly correlated with colour and stellar population than morphology. 
This would explain why no local population of galaxies selected by 
morphology displays the weak clustering of faint blue galaxies. 
This may also be the first detection of large population of galaxies at 
low redshift with similar clustering
properties to faint blue galaxies.  

\section{Summary}

We have used deep multicolour galaxy catalogues of two 
$5^\circ \times 5^\circ$ fields to study clustering from 
$z\sim 0$ to $z\sim 0.4$. The key conclusions of this paper are:
 
(i) The galaxy spatial correlation function is a power law on 
comoving scales less than $15 h^{-1} {\rmn Mpc}$. At larger scales, the
correlation function is consistent with a power law though a break
in the correlation function is not inconsistent with the data. 

(ii) Despite the large fields-of-view, 
there are significant differences in the measured amplitude 
of the clustering; with the possible exception of blue galaxies. It is clear 
that fields larger than $100 \Box ^\circ$ are required 
to accurately measure the clustering of $B_J\sim 22$ galaxies. 

(iii) Dwarf galaxies in relatively nearby clusters ($z\sim 0.11$) 
may effect estimates of faint galaxy correlation function. 
The effect is colour dependent with the clustering of red galaxies 
varying significantly between the 2 fields observed.

(iv) The clustering properties of galaxies strongly depend on the 
band used to select the catalogue. Bluer bands 
show weaker clustering than red bands and there is a rapid decline 
of the amplitude of the $B_J$ correlation function at faint magnitudes. 
It is probably inappropriate to fit a simple clustering model to
correlation functions derived from single band imaging due to 
the changing morphological mix with magnitude. 

(v) The clustering properties of galaxies strongly depend on colour. 
Such behaviour is consistent with colour being correlated with 
morphological type. Red galaxies (early types) exhibit 
stronger clustering with larger values of $\gamma$ 
than blue galaxies (late and irregular types). 
 
(vi) Blue galaxies have extremely weak clustering with 
$r_0\la 3 h^{-1} {\rmn Mpc}$. This is considerably weaker than 
the clustering of late type galaxies and is consistent 
with the clustering of galaxies being more strongly correlated with colour 
and stellar population than morphology. 

(vii) The clustering of $B_J<21.5$ blue galaxies is comparable to 
$B_J>23$ blue galaxies. This is strong evidence for star forming
galaxies being weakly clustered from $z\sim 0.4$ until the present
epoch. 

\section*{Acknowledgements}

The authors wish to thank the SuperCOSMOS unit at Royal Observatory 
Edinburgh for providing the digitised scans of UK Schmidt Plates
The authors also wish to thank Nigel Hambly, Bryn Jones 
and Harvey MacGillivray for 
productive discussions of the methods employed to coadd 
scans of photographic plates.
This research has made use of the NASA/IPAC Extragalactic Database 
which is operated by the Jet Propulsion
Laboratory, California Institute of Technology, under contract with the 
National Aeronautics and Space Administration. 
Michael Brown 
acknowledges the financial support of an Australian Postgraduate Award.

\label{lastpage}
\end{document}